# Stabilizing perpendicular magnetic anisotropy with strong exchange bias in PtMn/Co by magneto-ionics

Beatrice Bednarz,[1,a)] Maria-Andromachi Syskaki,[1, 2] Rohit Pachat,[3] Leon Prädel,[4] Martin Wortmann,[5] Timo Kuschel,[5] Shimpei Ono,[6] Mathias Kläui,[1] Liza Herrera Diez,[3] and Gerhard Jakob[1,a)]

[1)] *Institute of Physics, Johannes Gutenberg University Mainz, Staudingerweg 7, 55128 Mainz, Germany*
[2)] *Singulus Technologies AG, Hanauer Landstraße 103, 63796 Kahl am Main, Germany*
[3)] *Centre de Nanosciences et de Nanotechnologies, CNRS, Université Paris-Saclay, 10 Boulevard Thomas Gobert, 91120 Palaiseau, France*
[4)] *Max Planck Institute for Polymer Research, Ackermannweg 10, 55128 Mainz, Germany*
[5)] *Faculty of Physics, Bielefeld University, Universitätsstraße 25, 33615 Bielefeld, Germany*
[6)] *Central Research Institute of Electric Power Industry, Yokosuka, Kanagawa 240-0196, Japan*

[a)] Authors to whom correspondence should be addressed: bbednarz@uni-mainz.de, jakob@uni-mainz.de

**ABSTRACT**

Electric field control of magnetic properties offers a broad and promising toolbox for enabling ultra-low power electronics. A key challenge with high technological relevance is to master the interplay between the magnetic anisotropy of a ferromagnet and the exchange coupling to an adjacent antiferromagnet. Here, we demonstrate that magneto-ionic gating can be used to achieve a very stable out-of-plane (OOP) oriented magnetization with strong exchange bias in samples with as-deposited preferred in-plane (IP) magnetization. We show that the perpendicular interfacial anisotropy can be increased by more than a factor 2 in the stack Ta/Pt/PtMn/Co/HfO$_2$ by applying -2.5 V gate voltage over 3 nm HfO$_2$, causing a reorientation of the magnetization from IP to OOP with a strong OOP exchange bias of more than 50 mT. Comparing two thicknesses of PtMn, we identify a notable trade-off: while thicker PtMn yields a significantly larger exchange bias, it also results in a slower response to ionic liquid gating within the accessible gate voltage window. These results pave the way for post-deposition electrical tailoring of magnetic anisotropy and exchange bias in samples requiring significant exchange bias.

In view of the climate crisis and the rapidly growing power consumption of information and communication technology, more energy-efficient data storage and processing is needed.[1,2] In that respect, voltage control of magnetic properties is one of the central approaches with high potential for ultralow-power data storage as well as energy-efficient hardware for dissipative and neuromorphic computing.[3–8] Particularly large magneto-electric effects of over 7000 fJ/Vm, calculated as the change in magnetic anisotropy divided by the applied electric field, can be obtained with magneto-ionics using the electric field induced motion of mobile ions.[9] Magneto-ionics was shown to allow for low-power switching of magnetization,[10] manipulation of the exchange bias (EB)[11,12] and the Dzyaloshinskii–Moriya interaction[13] as well as the control of domain wall nucleation and motion.[14,15]

The EB is fundamental to many technological applications. It describes the pinning of the magnetization of a ferromagnetic (FM) material by the uncompensated magnetic moments of an adjacent antiferromagnet (AFM), causing a shift of the magnetization curve along the magnetic field direction. Its applications range from pinning the reference layer in spin valves for magnetic memories, sensors, and spintronic devices[16], to replacing the auxiliary magnetic field and allowing for field-free switching of magnetization[17], to imprinting domain patterns for magnetophoretic applications.[18] So far, research on the control of the EB focused on the manipulation of the EB in samples with a fixed magnetization direction.[5,11,12,19,20] This was achieved by transfer of different ions, e.g. O,[21] H,[12] Li,[22] as well as N.[23] Different material systems were shown to allow for the manipulation of the EB ranging from systems with the AFM on top[19] or below[11] the FM and with insulating[11] as well as metallic AFMs.[19] However, so far, a demonstration of the possibility to rotate the magnetization from an in-plane (IP) to an out-of-plane (OOP) state while retaining a strong EB is still missing. This would be highly desirable e.g. to program 3D magnetization sensors for which one part of the sensor needs to have IP magnetization and another part OOP magnetization.

In this paper, we investigated the effect of magneto-ionic gating on the EB between platinum-manganese (PtMn) and cobalt (Co) thin films. We applied the ionic liquid gating technique, enabling large-area gating with simple sample fabrication and applicability to a wide range of materials. In the as-grown state the system exhibits IP magnetization. By gating, we demonstrate the rotation of the magnetization into a non-volatile state with OOP magnetization and OOP EB. Notably, this OOP state is the energetically favorable one once it is reached. To better understand the influence of the AFM layer thickness, we compare two stacks with different PtMn thicknesses. We show that there is a trade-off between the strength of the EB in the final state and the speed of the gating process within the accessible gate voltage window until substantial oxidation of the Co. In thinner (8 nm) samples the magnetization can be rotated faster while thicker (20 nm) samples show a stronger EB.

We have studied magnetic stacks with the composition Ta (5 nm)/Pt (3 nm)/PtMn ($t$)/Co (0.9 nm)/HfO$_2$ (3 nm) deposited on thermally oxidized Si/SiO$_2$ substrates, with two thicknesses of PtMn ($t$ = 8 and 20 nm) (Fig. 1a). Co was chosen as the FM material because it can have strong perpendicular magnetic anisotropy (PMA). PtMn, a metallic AFM, is commonly used in the industry because of its ability to provide a strong EB with high thermal stability.[24,25] The seed layers Ta and Pt provide a smooth surface and matching lattice constant for an optimized growth of PtMn. Co was capped with HfO$_2$, which serves as a donor for mobile oxygen species. It is a high-κ dielectric with large technological relevance[26] and allows for high ion-mobility at low voltages. To ensure very high sample quality, the stacks were sputtered in an industrial Singulus Rotaris magnetron sputtering tool at room temperature. As an isolating layer, HfO$_2$ was sputtered using RF voltages. The further layers of the stack are metallic and sputtered by DC voltages. To move the mobile oxygen species inside the HfO$_2$, a droplet of the ionic liquid 1-ethyl-3-methylimidazolium-bis(trifluoromethylsulfonyl)-imide ([EMIM]$^+$ [TFSI]$^-$) was added on the sample surface, and a glass plate coated with indium tin oxide (ITO) was placed on top. The gate voltage was then applied from the ITO electrode across the ionic liquid and HfO$_2$ to the metallic layers at the bottom of the material stack, which were contacted by wire bonding (Fig. 1a). In this way, a sample area of approximately 5x5 mm$^2$ was gated.

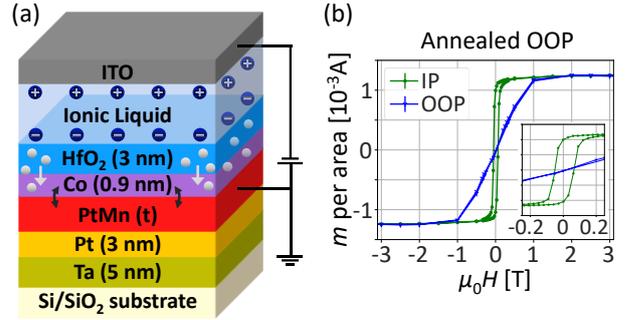

**FIG. 1.** (a) Sample stack. The white spheres represent the oxygen ions which can be moved by applying a gate voltage across the HfO$_2$. The arrow between the Co and the PtMn represents the EB coupling between the two layers. (b) SQUID measurements of the sample with 8 nm PtMn after annealing in an OOP magnetic field of 800 mT for 2 h at 300 °C. The SQUID measurements were performed along the IP (green) and OOP (blue) sample direction to confirm the fully IP magnetization of the Co. The inset shows a zoom into the center of the SQUID loops.

In the as-deposited state, the samples show full IP magnetization for both 8 nm and 20 nm PtMn (SM Fig. S3). We aim to explore whether it is possible to reach OOP magnetization with OOP EB through ionic-liquid gating. Therefore, the samples were annealed in an OOP magnetic field of 800 mT at 300 °C for 2 h, to establish OOP EB. Annealing the samples leaves the magnetization IP, while the coercivity of the IP magnetization curve increases significantly (Fig. 1b). The lateral crystal orientation of PtMn (111) is not visibly affected by the annealing, as confirmed by x-ray diffraction (SM Fig. S1).

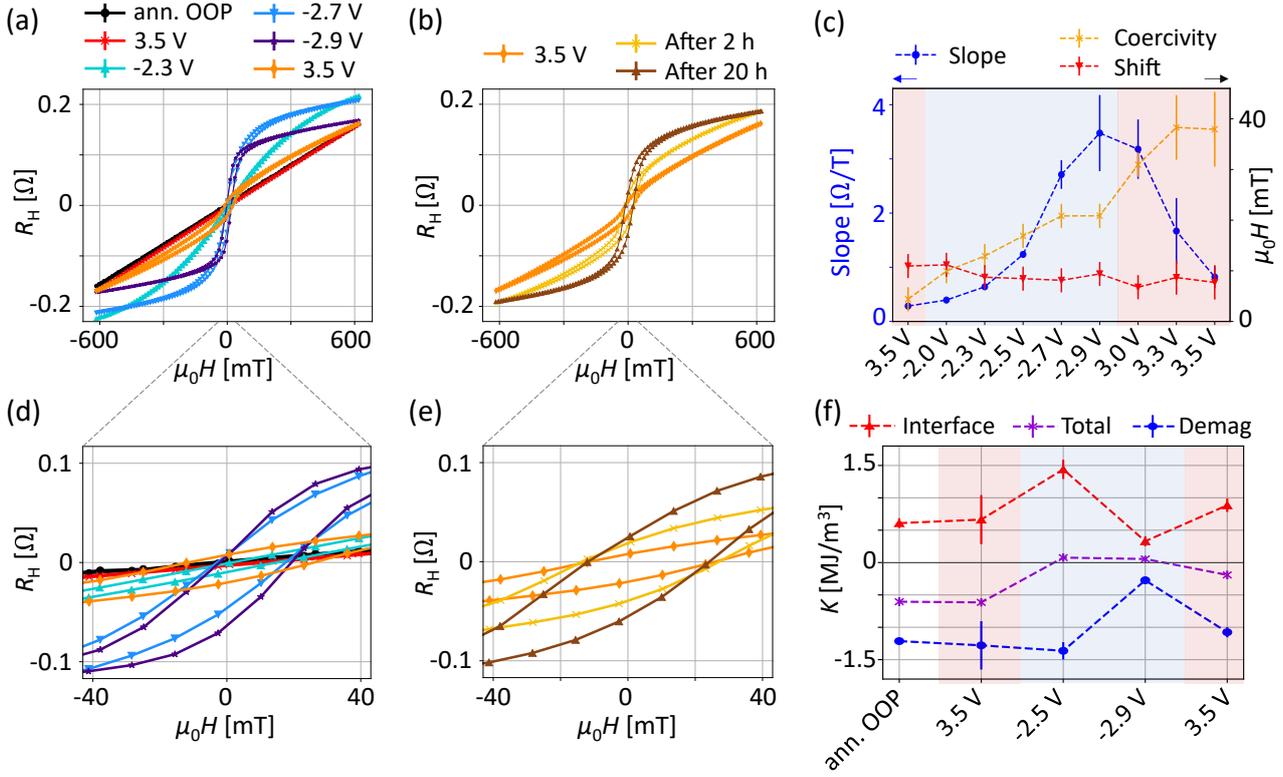

**FIG. 2.** Evolution of magnetic properties by gating in the sample with 8 nm PtMn. (a) AHE curves from the first gating cycle. The initialization gate voltage of 3.5 V was applied for 5 min. All further gate voltages were applied for 2 min. (b) Time evolution of the AHE curve after the magnetization was brought back toward an IP state by applying a gate voltage of 3.5 V. (c) Overview over slope, coercivity and EB shift of the AHE loops at zero remanent magnetization. (d) and (e) Zoom into subfigures (a) and (b). (f) Evolution of the total anisotropy, demagnetization, and interfacial energy per volume, obtained from SQUID measurements at 300 K (SM Fig. 3). The initialization voltage of 3.5 V was applied for 10 min, all further voltages for 5 min each.

The gate voltages were applied using a Keithley 2400. For each gating step, the voltage was slowly increased to minimize instabilities until reaching the desired voltage. After waiting for the desired gating time, the gate voltage was turned off and the magnetization curve was measured via the anomalous Hall effect (AHE) at room temperature. The ordinary Hall component was subtracted for each curve (see supplementary material SM2). We use the slope of the anomalous Hall resistance curve at zero remanent magnetization as a qualitative measure of the extent to which the magnetization curve is dominated by magnetic easy- or hard-axis behavior.[27]

We will first discuss the samples with the thinner PtMn layer of 8 nm. To establish a defined initial state, first, a positive gate voltage of 3.5 V was applied. This positive voltage induces an upward drag on the oxygen species inside the $HfO_2$ layer. The Co is set to a state of minimum oxygen concentration. As shown in Fig. 2a, the magnetization remains IP after this initialization step, with only minor changes in the magnetization curve. Subsequently, negative voltages were applied. In agreement with previous reports[28], we observe a progressive transformation toward a square-shaped magnetization loop. Since the AHE probes the OOP magnetization, this demonstrates that the magnetization rotates from the IP to the OOP direction. All AHE curves are shifted toward positive magnetic field values revealing the OOP EB (Fig. 2d).

The OOP magnetization, which is reached by applying negative gate voltages, is very stable over time (SM Fig. S5). By applying positive gate voltages, the magnetization can be rotated back toward the IP direction. However, this IP state is no longer stable over time. Within a few hours, the sample evolves back to the OOP state (Fig. 2b). Additionally, it is not possible in this material system to bring the sample to an IP state by overoxidation. Instead, at -2.9 V, we observe a decrease in the total Hall resistance, while the slope continues to increase. This demonstrates the stabilization of the OOP state with EB in an initially IP sample by ionic liquid gating.

Fig. 2c shows the evolution of the EB shift and coercivity in comparison to the slope of the AHE loop during the different gating steps. The coercivity of the AHE loops increases when the magnetization rotates from IP toward OOP. When positive voltages are applied subsequently, the magnetization direction is no longer stable in time, and the AHE curve is drifting during the measurement. The resulting coercivities therefore have a large error. The EB shift initially decreases slightly when a negative voltage is applied. Then, it stabilizes and reaches a relatively constant value of 8.2 mT with a standard deviation of 0.8 mT and a systematical error, indicated by the error bars, of 2.4 mT (see SM7). This stabilization of the EB shift was observed in several nominally identical samples (SM Fig. S8). In the annealed state, the EB shift fluctuates. However, by performing the first gating cycle, it stabilizes for all nominally identical samples to an identical value within the error margin. This indicates a varying amount of oxygen ions at the PtMn/Co interface in the annealed state, which are removed by the positive gate voltage.

To better understand the underlying magnetic changes in this system, we performed magnetization measurements in a superconducting quantum interference device (SQUID) (SM Fig. 3). From the IP and OOP SQUID loops, we obtained the magnetic anisotropy energies per volume after different gating steps (Fig. 2f). In this material system, we consider the demagnetization energy, which favors IP magnetization, and the interfacial anisotropy energy, which favor OOP magnetization, as the most important energy contributions. The total anisotropy energy was obtained from the area between the OOP and IP magnetization curves (Eq. 1). The demagnetization energy was calculated using its proportionality to the squared saturation magnetization $M_S^2$ (Eq. 2).[29] The interfacial anisotropy energy, stemming from the two Co interfaces, can then be calculated using Eq. 3. The magnetization $M$ and all anisotropies were calculated using the nominal thickness of Co.

$$K_{\text{total}} = \frac{1}{\mu_0} \int_{shift}^{\mu_0 H_{max}} (\mu_0 M_{OOP} - \mu_0 M_{IP}) d(\mu_0 H) \quad (1)$$

$$K_{\text{demag}} = \frac{1}{2} \mu_0 M_S^2 \quad (2)$$

$$K_{\text{interface}} = K_{total} - K_{demag} \quad (3)$$

In Fig. 2f, the total anisotropy energy (purple) shows a clear trend of becoming more negative (toward IP magnetization) when applying positive voltages and becoming more positive (toward OOP magnetization) when applying negative voltages. This supports the presumed underlying mechanism of oxygen ion migration. The total anisotropy energy reaches its maximum positive value (maximum PMA) at -2.5 V. This is explained by the strong increase of the interfacial anisotropy (red), of a factor of 2.2, after the application of -2.5 V. This, in turn, can be explained by the additional oxygen at the Co interfaces, especially the Co/HfO$_2$ one. It allows for a stronger hybridization between the Co-3$d$ and the O-2$p$ orbitals, which is known to cause interfacial PMA.[30] Notably, the demagnetization energy (blue) remains quite constant and even slightly increases up to -2.5 V, in accordance with the observations from the AHE effect measurements. This demonstrates that the rotation of the magnetization is caused by the increasing interfacial PMA and not by a reduction of the demagnetization energy.

After the application of -2.9 V, both the demagnetization and the interfacial anisotropy energy drop significantly. For the demagnetization energy, this is clearly related to the reduction of the total magnetic moment caused by the overoxidation of Co. The influence of the overoxidation can be described as a change in the average magnetization or in the Co layer thickness, which, however, does not alter the qualitative results (SM Fig. 4). For the interfacial anisotropy energy, its reduction due to overoxidation suggests that the oxygen content at the Co/HfO$_2$ interface has gone past the optimum oxidation point which corresponds to maximum PMA or that the PMA at the PtMn/Co interface got diminished. This interfacial PMA cannot be fully recovered by applying a positive voltage of 3.5 V.

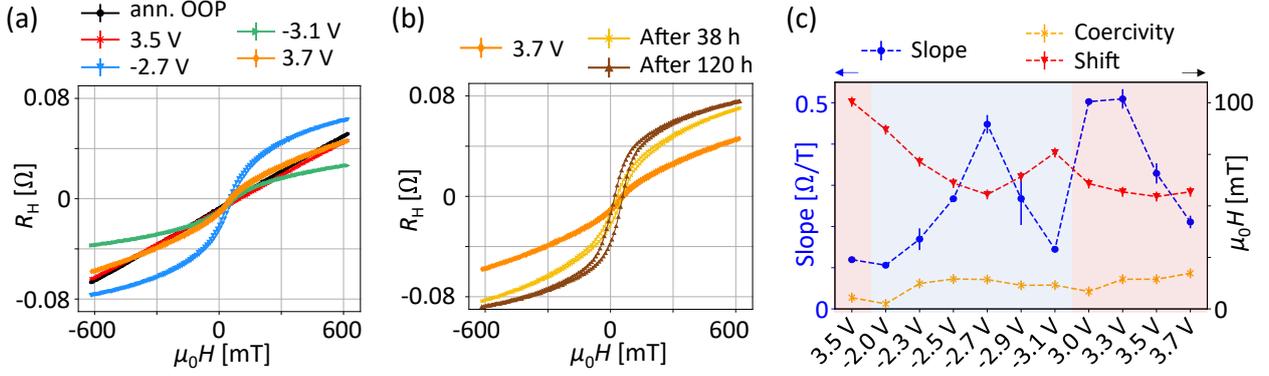

**FIG. 3.** *Evolution of magnetic properties by gating in the sample with 20 nm PtMn. (a) AHE curves from the first gating cycle. All gate voltages were applied for 2 min. A larger version of this subfigure which also includes the curves at -2.3 and -2.9 V can be found in SM6. (b) Time evolution of the AHE loop after the magnetization was brought back toward an IP state by applying 3.7 V. (c) Overview over slope, coercivity and EB shift of the AHE loops.*

From the results on the AHE measurements, we can, however, assume that over time the oxygen ions will diffuse back to the positions which lead to the maximum PMA.

So far, we have shown that we can establish a stable OOP magnetization with OOP EB in a sample with initial IP magnetization by ionic-liquid gating. However, for applications, a much larger EB is required. Therefore, we also investigated a sample with thicker PtMn of 20 nm (Fig. 3). Equivalently to the stack with 8 nm PtMn, the magnetization of the sample is initially IP and stays IP after applying +3.5 V (Fig. 3a). As expected, the AHE loop evolves toward a square-shaped magnetization loop by negative gating. Like in the sample with thinner PtMn, applying -2.9 V or larger negative voltages leads to overoxidation of the Co, expressed by a decrease in the total Hall resistance. However, the coercivity and the slope remain low for the applied gate voltages in the sample with thicker PtMn (Fig. 3c). This changes only after performing the full first gating cycle and then waiting five days (Fig. 3b). Most likely, this gives the oxygen ions time to diffuse into the energetically favored lattice positions where they can hybridize with the Co atoms, resulting in increased interfacial PMA. Presumably, this slower reorientation of the magnetization towards the OOP state, both after applying the accessible negative voltages as well as by waiting after overoxidation, is caused by a higher degree of crystallinity of Co. The Co is grown on thicker and therefore more crystalline PtMn and a higher degree of crystallinity is known to slow down oxygen diffusion.[31] This highlights the importance of the underlayer on the gating dynamics.

Just like in the thinner sample, the OOP state becomes energetically favorable as soon as it is reached once. A positive voltage initially brings the magnetization from the OOP state back toward an IP state, which however evolves back toward the OOP direction over time (Fig. 3b). Notably, the EB shift stabilizes to a value above 50 mT in this sample, which is more than 5 times greater than in the thinner sample.

To verify whether the magnetic changes are caused by an oxidation of Co, we performed x-ray photoelectron spectroscopy (XPS) measurements. Fig. 4a shows the XPS spectra at the Co 2p-edge in the annealed state (top) and after applying a gate voltage of -2.7 V (bottom). The annealed state shows a pure metallic $Co^0$ (blue peaks). By gating, a clear $Co^{+II}$ signal arises (orange peaks), confirming the partial oxidation of the Co. From SQUID measurements we know that for the oxidized sample, around half of the magnetization is lost in the oxidized state. The strong reduction of the $Co^0$ peak intensity in the oxidized sample is most likely due to the surface sensitivity of XPS, indicating a stronger oxidation in the upper layers of the Co compared to the lower ones.

Understanding the oxidation of the interface between Co and PtMn is very important to retain a strong EB. Therefore, its oxidation was probed by measuring XPS at the Mn 2p-edge (Fig. 4b). Overall, the signal-to-noise ratio is worse, since the Mn atoms are lying further down below the sample surface. In the annealed state, the sample consists mainly of metallic $Mn^0$ (green peaks). However, after applying the negative gate voltage, also the Mn shows a clear signal of oxidation (magenta peaks). Some oxygen therefore reached the PtMn/Co interface. This is likely the reason why the EB initially drops with negative gating. However, the AHE data revealed that the EB stabilizes at a value larger than zero. This indicates that the interface does not fully get oxidized or that this oxidation does not break the exchange coupling completely. Additionally, the thinning of the Co layer by oxidation reduces the Zeeman energy and consequently leads to a relative increase in the exchange bias field. Overall, this allows for the retention of a significant EB.

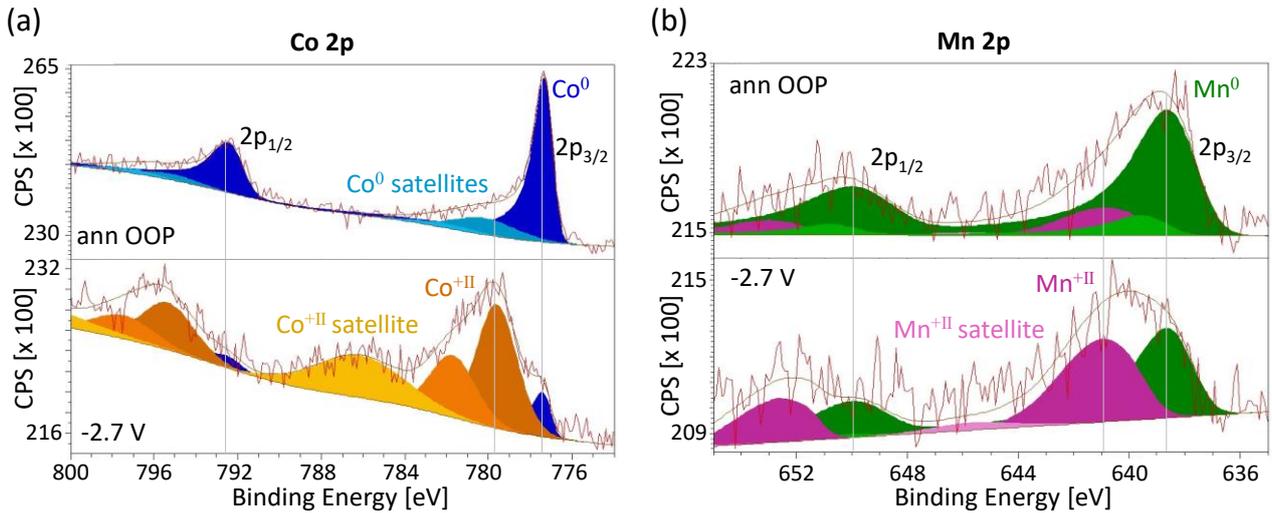

*FIG. 4. XPS spectra of the sample with 8 nm PtMn, measured at (a) the Co 2p-edge and (b) the Mn 2p-edge in the states after the annealing in an OOP magnetic field and after applying -2.7 V for 5 min, respectively. The $Co^{+II}$ and $Mn^0$ peaks show a multiplet splitting. Details on the experiment and the fitting procedure are given in supplementary material SM9.*

In summary, we achieved the stabilization of a magnetic state with OOP magnetization and OOP EB in a sample with initial IP magnetization by ionic liquid gating. In the stack Ta (5 nm)/Pt (3 nm)/PtMn (8 nm)/Co (0.9 nm)/$HfO_2$ (3 nm), we showed that by applying a gate voltage of -2.7 V the magnetization could be reoriented to OOP, with an exchange bias shift of (8.2 ± 2.4) mT. SQUID and XPS measurements confirmed that the changes are due to oxygen migration in the sample. As the oxygen ions reach the Co/$HfO_2$ interface, the interfacial anisotropy increases, causing the reorientation of the magnetization towards the OOP direction. At large negative voltages of more than -2.7 V both the demagnetization energy and the interfacial anisotropy energy decrease due to overoxidation of the Co. Subsequent positive voltages lead to a reorientation of the magnetization back toward the initial IP state. However, this state is volatile and evolves back toward the OOP state over time, demonstrating that the OOP state becomes the energetically favorable state once it is reached. Some of the oxygen ions also reach the bottom interface of the Co after negative gating, visible in a partial oxidation of the Mn atoms in PtMn. Notably, this oxidation does not lead to a loss of EB, just to an initial reduction of EB which then saturates.

The sample with the thicker PtMn film of 20 nm exhibits a slower gating dynamics in the accessible voltage window, highlighting the importance of the PtMn/Co interface on the gating process. Notably, in the final state, we found fully shifted magnetization loops with a strong EB of more than 50 mT.

These results contribute to the advancement of easier and better device fabrication for energy-efficient data storage and processing as well as sensor applications.

## SUPPLEMENTARY MATERIAL (SM)

See the supplementary material for more detailed information on measurements and data analysis.

## ACKNOWLEDGEMENTS

This project has received funding from the European Union's Horizon 2020 Research and Innovation Programme under the Marie Skłodowska-Curie grant agreement No 860060 "Magnetism and the effect of Electric Field" (MagnEFi), as well as from the Deutsche Forschungsgemeinschaft (DFG, German Research Foundation) – TRR 173/2-#268565370 Spin+X (Projects A01 and B02). This work was partly founded by JSPS KAKENHI Grand Number JP21H05016. We thank the Max Planck Institute for Polymer Research for access to their XPS tool.

## AUTHOR DECLARATIONS

**Conflict of interest**
The authors have no conflicts to disclose.

## DATA AVAILABILITY

Data that support the findings of this study are openly available in Zenodo at
https://doi.org/10.5281/zenodo.11395888.[32]

# Supplementary Material (SM)

## Stabilizing perpendicular magnetic anisotropy with strong exchange bias in PtMn/Co by magneto-ionics

Beatrice Bednarz,[1,a)] Maria-Andromachi Syskaki,[1, 2] Rohit Pachat,[3] Leon Prädel,[4] Martin Wortmann,[5] Timo Kuschel,[5] Shimpei Ono,[6] Mathias Kläui,[1] Liza Herrera Diez,[3] and Gerhard Jakob[1,a)]

[1)] *Institute of Physics, Johannes Gutenberg University Mainz, Staudingerweg 7, 55128 Mainz, Germany*
[2)] *Singulus Technologies AG, Hanauer Landstraße 103, 63796 Kahl am Main, Germany*
[3)] *Centre de Nanosciences et de Nanotechnologies, CNRS, Université Paris-Saclay, 10 Boulevard Thomas Gobert, 91120 Palaiseau, France*
[4)] *Max Planck Institute for Polymer Research, Ackermannweg 10, 55128 Mainz, Germany*
[5)] *Faculty of Physics, Bielefeld University, Universitätsstraße 25, 33615 Bielefeld, Germany*
[6)] *Central Research Institute of Electric Power Industry, Yokosuka, Kanagawa 240-0196, Japan*

[a)] Authors to whom correspondence should be addressed: bbednarz@uni-mainz.de, jakob@uni-mainz.de

## SM1. X-ray diffraction (XRD) measurements to examine the crystallographic structure

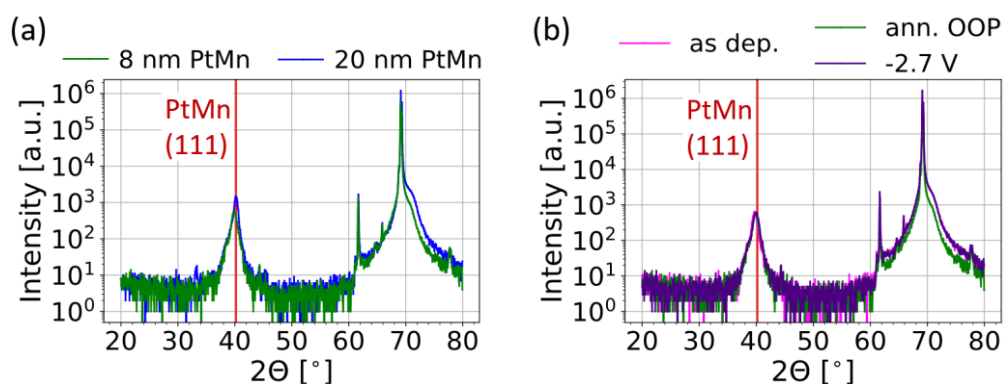

**FIG. S1.** XRD patterns of the samples Ta (5 nm)/Pt (3 nm)/PtMn (t)/Co (0.9 nm)/HfO$_2$ (3 nm) (a) comparing samples with the thicknesses t = 8 nm and 20 nm PtMn after annealing in an out-of-plane (OOP) magnetic field of 800 mT for 2 h at 300°C. (b) Influence of the annealing in the OOP magnetic field and subsequent gating with -2.7 V on the crystal structure for the sample with t = 8 nm PtMn. The red line in both figures shows the reference position of 40.2° for PtMn(111)[1]. The peak at 69.2° is the peak of the Si substrate. The sharp lines at 61.7° and 65.9° are resulting from the substrate and residual contributions of copper $K_\beta$ and tungsten $L_\alpha$ radiation to the incoming beam.

The structural characterization of the samples was done by XRD, measured in a Bruker D8 X-ray diffractometer. All spectra show the PtMn (111) peak at 40.2° with a shoulder on its left, corresponding to the Pt (111) peak. Fig. S1a shows that in both samples, with 8 nm as well as 20 nm PtMn, the PtMn grows in (111) orientation. The PtMn peak in the sample with 20 nm PtMn is larger and the center matches better with the theory position of the PtMn (111) peak, indicating less strain in the sample. Fig. S1b shows that the annealing in an OOP magnetic field of 800 mT for 2 h at 300°C, as well as gating at -2.7 V does not visibly influence the crystal structure of the PtMn. The peak at 69.2° corresponds to the Si (004) substrate peak. The peaks at its left edge are the corresponding satellite peaks, observed because the beam monochromatization by the Goebel mirror is incomplete.

## SM2. Hall measurements in a superconducting magnet cryostat to determine the OHE

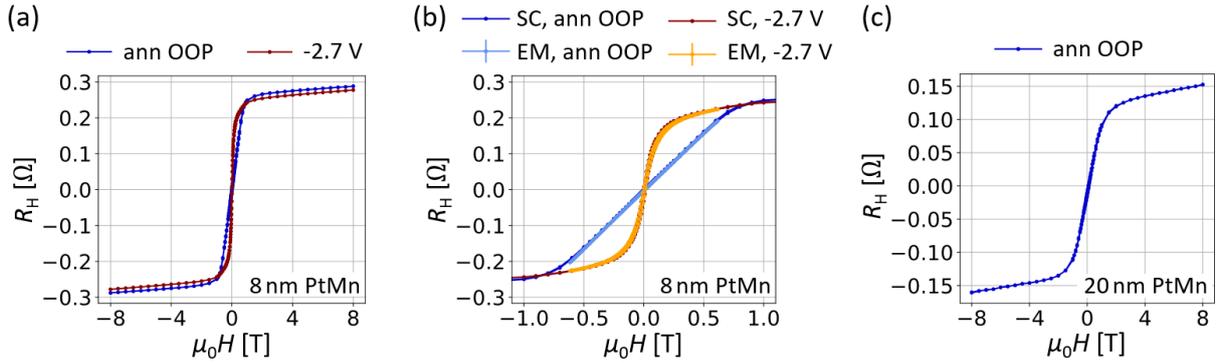

**FIG. S2.** *Hall measurements in a superconducting magnet cryostat, measured to determine the ordinary Hall effect (OHE) from the high field slope. From (a) a slope of m = $(3.3 \pm 0.3) \cdot 10^{-3} \Omega/T$ was obtained for the sample with 8 nm PtMn after saturation. From (c) a slope of m = $(3.9 \pm 0.3) \cdot 10^{-3} \Omega/T$ was obtained for the sample with 20 nm PtMn. (b) To confirm, that the magnetization curve measured in the superconducting magnet cryostat (SC) is identical to the one from the electromagnet (EM), the loops were plotted on top of each other.*

The anomalous Hall effect (AHE) curves discussed in the main text were measured at room temperature in an electromagnet. As the magnetic field strength, which can be reached with the electromagnet, is not large enough to saturate all magnetization curves, the ordinary Hall component was estimated from AHE measurements using a superconducting magnet (Fig. S2). The resulting ordinary Hall component was subtracted from all AHE measurements obtained with the electromagnet.

To estimate the ordinary Hall effect (OHE), magnetization curves were obtained by measuring the transverse resistivity with an applied current of 1 mA in a superconducting magnet cryostat at *T*=290 K and up to magnetic field values of 8 T. The samples were connected by wire bonding the corners of the unpatterned samples (size of 5 mm × 10 mm). 2 ms long pulses of +1 mA and -1 mA were applied with a Keithley 6221 current source along the sample diagonal. The transverse Hall voltage was measured with a Keithley 2182A nanovoltmeter in Delta mode and was averaged out for 100 data points to remove noise due to temperature fluctuation and thermoelectric contributions. For every magnetic field value this procedure was repeated, and the Hall voltage measured for currents applied along both diagonals of the sample ($R_{\text{Hall,dir1}}$ and $R_{\text{Hall,dir2}}$). The van der Pauw-method was used to eliminate the longitudinal component in the Hall signal. For that purpose, the following formula was used at every field step:

$$R_{\text{Hall,corr}} = \frac{R_{\text{Hall,dir1}} + R_{\text{Hall,dir2}}}{2} \tag{S1}$$

The resulting magnetization curves are plotted in Fig. S2. The slope was determined by fitting the data between 8 T and 4.5 T, as well as between -8 T and -4.5 T. For sample MASA697 with 8 nm PtMn, two samples were measured, one after the initial annealing and one after applying -2.7 V. In this case, the obtained slopes of the two samples were averaged, and the error was calculated as the standard deviation. For sample MASA699 with 20 nm PtMn, only one sample was measured, so only the two slopes at positive and negative magnetic field values were averaged. The error was determined as half of the difference between the two slopes.

Figure S2a shows that the slopes obtained for the sample with 8 nm PtMn, in the annealed state and after applying -2.7 V, are very similar, so assuming a constant OHE for all gating steps seems reasonable. Additionally, Figure S2b shows that the same Hall resistances were obtained with the superconducting magnet cryostat (SC) and the electromagnet (EM). Therefore, the slopes obtained from the SC measurements at high fields were subtracted from all measurements performed with the EM. For time and capacity reasons, it was not possible to perform all measurements in the superconducting magnet cryostat.

## SM3. SQUID measurements to characterize the magnetization in different states

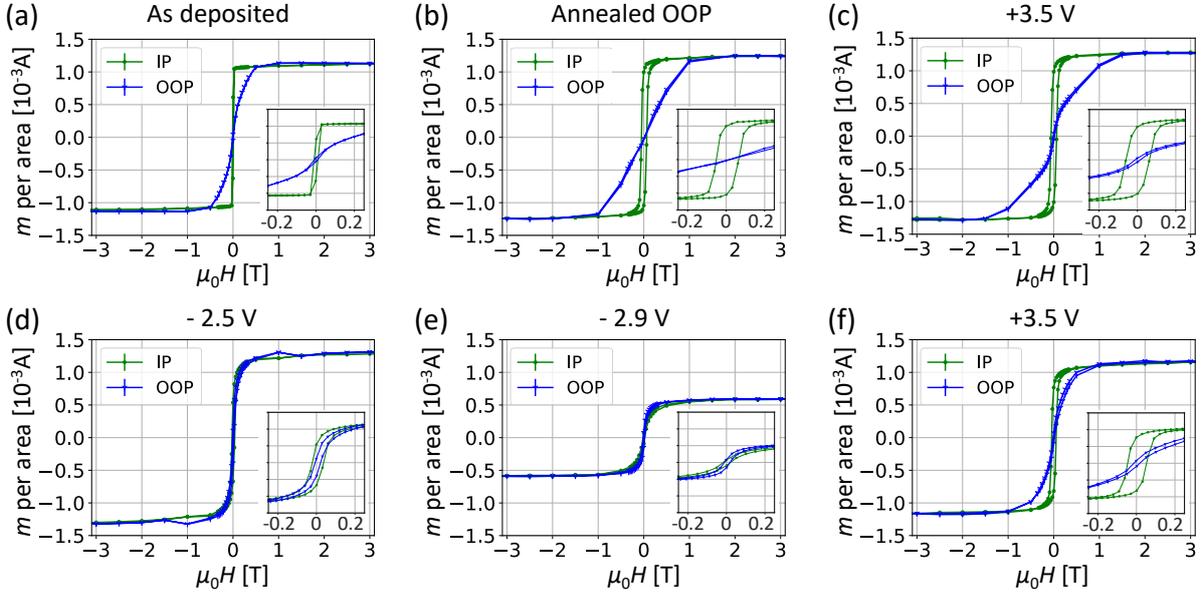

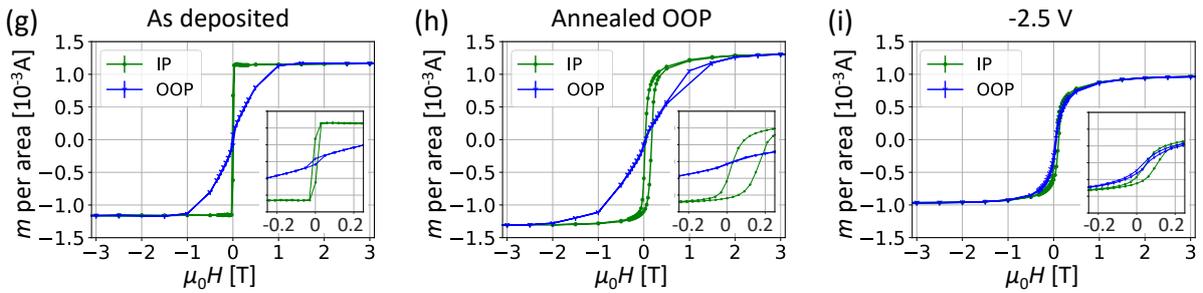

*FIG. S3.* IP and OOP SQUID measurements of the samples Ta (5 nm)/Pt (3 nm)/PtMn (t)/Co (0.9 nm)/HfO$_2$ (3 nm), with 8 nm (a-f) and 20 nm PtMn (g-i), respectively. The measurements were done in the as-deposited state, after annealing in an OOP magnetic field of 800 mT for 2 h at 300°C and after different gating steps. The first positive initialization voltage was applied for 10 minutes, all other voltages for 5 minutes, respectively. The insets show the zoom into the center of the magnetization curves. All curves were measured up to magnetic fields of ±4 T.

Superconducting quantum interference device (SQUID) measurements were performed using the Quantum Design MPMS-XL5 SQUID Magnetometer to analyze the initial magnetic state and the changes due to gating. Samples with both 8 and 20 nm PtMn were investigated. Before mounting the samples in the SQUID, the ITO-coated glass plate was removed, the ionic liquid washed off with acetone and the ungated areas of the sample were cut off.

Initially, before applying negative gate voltages, both samples show an in-plane (IP) magnetization (Fig. S3a-c and g-h). The small jump at the center of the OOP magnetization loop of the sample with 8 nm PtMn after the application of 3.5 V indicates a small canting in the magnetization (Fig. S3c). After negative gating of -2.5 V, a remanent moment is observed in both IP as well as OOP direction (Fig. S3d and i). When applying -2.9 V, the magnetic moment per area decreases (Fig. S3e), as was also observed in the AHE loop. The OOP magnetization loop remains a magnetic easy axis loop, while the IP magnetization loop changes slightly towards a magnetic hard axis loop. Subsequent gating with a positive voltage of 3.5 V brings the magnetization back towards the IP direction (Fig. S3f).

For each magnetization curve, the data points with low fit quality <0.7 were removed. The deviations from an ideal point dipole were corrected with a geometry factor, and a small offset in the magnetic field values caused by magnetic field lines trapped in the superconducting magnet was corrected. Then the diamagnetic contribution was subtracted from the data by subtracting the slope obtained from the first 3 data points.

## SM4. Anisotropies calculated from the SQUID data

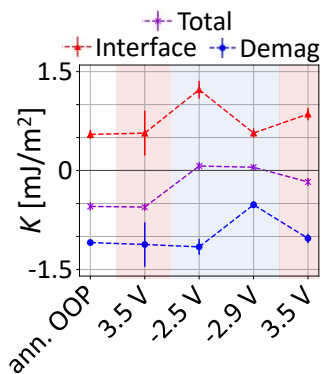

**FIG. S4.** *Evolution of the total anisotropy, demagnetization and interfacial energy per area, obtained from SQUID measurements, in the sample with 8 nm PtMn. The measurements were taken at 300 K and the energies were calculated under the assumption of constant magnetization and changing Co thickness (SM Fig. 3). The initialization voltage of 3.5 V was applied for 10 minutes, all further voltages for 5 minutes each.*

To investigate the influence of the gating on the magnetic anisotropy, the changes in the total anisotropy energy, demagnetization energy and interfacial anisotropy energy were examined. To calculate the demagnetization energy from the SQUID loops the saturation magnetization $M_S$ is required. It is defined as the total magnetic moment per magnetic volume. However, from the XPS measurements we know that the upper layers of the Co are getting oxidized by gating, likely forming an oxygen gradient inside the sample. Therefore, the two limiting cases have to be considered by calculating the average magnetization in the Co layer using the nominal Co thickness, as well as by assuming that the magnetization remains constant, and oxidation purely reduces the Co thickness. In Fig. 2f in the main text, the changes of anisotropy with gating were shown for the case of fixed Co thickness and changing average magnetization. In Fig. S4 the second case is shown, assuming a constant magnetization. Both figures show the same qualitative behavior.

The anisotropies were calculated from the SQUID loops shown in Fig. S3. To obtain the total anisotropy energy, the integrals of the IP and OOP magnetization curves were calculated between the center of the curve, corresponding to its shift, and 2 T. The integrals from the positive and negative side of the curve were averaged and the difference used as the error. Then, the obtained integral from the IP magnetization curve was subtracted from the integral of the OOP curve.

The error on the total anisotropy energy was calculated by error propagation, considering the errors on the integrals of the IP and OOP magnetization curves, as well as the error of the data points close to the zero crossing of the magnetization. This last error was estimated using the sample with 8 nm PtMn after applying -2.5 V (Fig. S3d). In that sample, the zero crossing of the magnetization was very close to ±1 T. By leaving out these datapoints and calculation the change of the total anisotropy energy, an error of 0.05 MJ/m$^3$ was obtained.

The error on the demagnetization energy was found by error propagation from the error on $M_s t$. For some of the measurements with a magnetic hard axis loop in the OOP direction, the obtained values for the maximum magnetic moment per area differed between the corresponding IP and OOP measurements. In these cases, the $M_s t$ from the IP magnetization curves was chosen because of the better data quality and reliability of the easy axis magnetization curves. For the error on $M_s t$, half of the difference between $M_s t$ obtained from the IP and OOP SQUID loops was used.

For the interfacial anisotropy energy, the error was obtained by error propagation considering the errors on the demagnetization and the total anisotropy energy.

## SM 5: Stability of the state with OOP magnetization and AHE measurement procedure

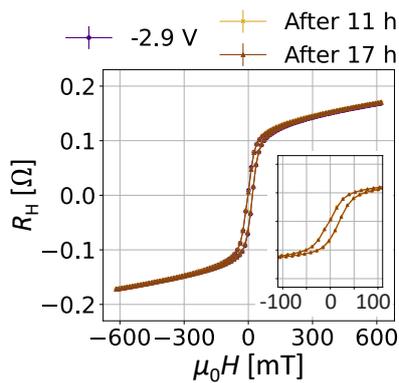

**FIG. S5.** *Stability check of the OOP state of the sample Ta (5 nm)/Pt (3 nm)/PtMn (8)/Co (0.9 nm)/HfO$_2$ (3 nm) after applying -2.9 V (1$^{st}$ gating cycle), measured via the AHE.*

Fig. S5 shows the stability of the magnetic state after the application of -2.9 V. No change is visible over 17 h waiting time. The AHE data was measured at room temperature in an electromagnet. For all measurements in the electromagnet, we used the lock-in technique at 625 Hz with the tensormeter RTM2-501 from the HZDR Innovation GmbH. The datapoints at every field step were averaged and the OHE, obtained from the measurements in the superconducting magnet cryostat, subtracted. The error bars, obtained from averaging the data points at every magnetic field step, are smaller than the marker size.

## SM 6: Enlarged images of the AHE data from the 1$^{st}$ gating cycle

For improved visibility, the Fig. 2a (sample with 8 nm PtMn) and 3a (sample with 20 nm PtMn) from the main manuscript are shown here in an enlarged version. For the sample with 20 nm PtMn, only a subset of the magnetization curves was shown in the main manuscript to enhance the clarity of the figure. Here, the data for -2.3 V and -2.9 V gate voltage is additionally added. It shows the same trend for the evolution of the squareness as the respective data from the sample with 8 nm PtMn.

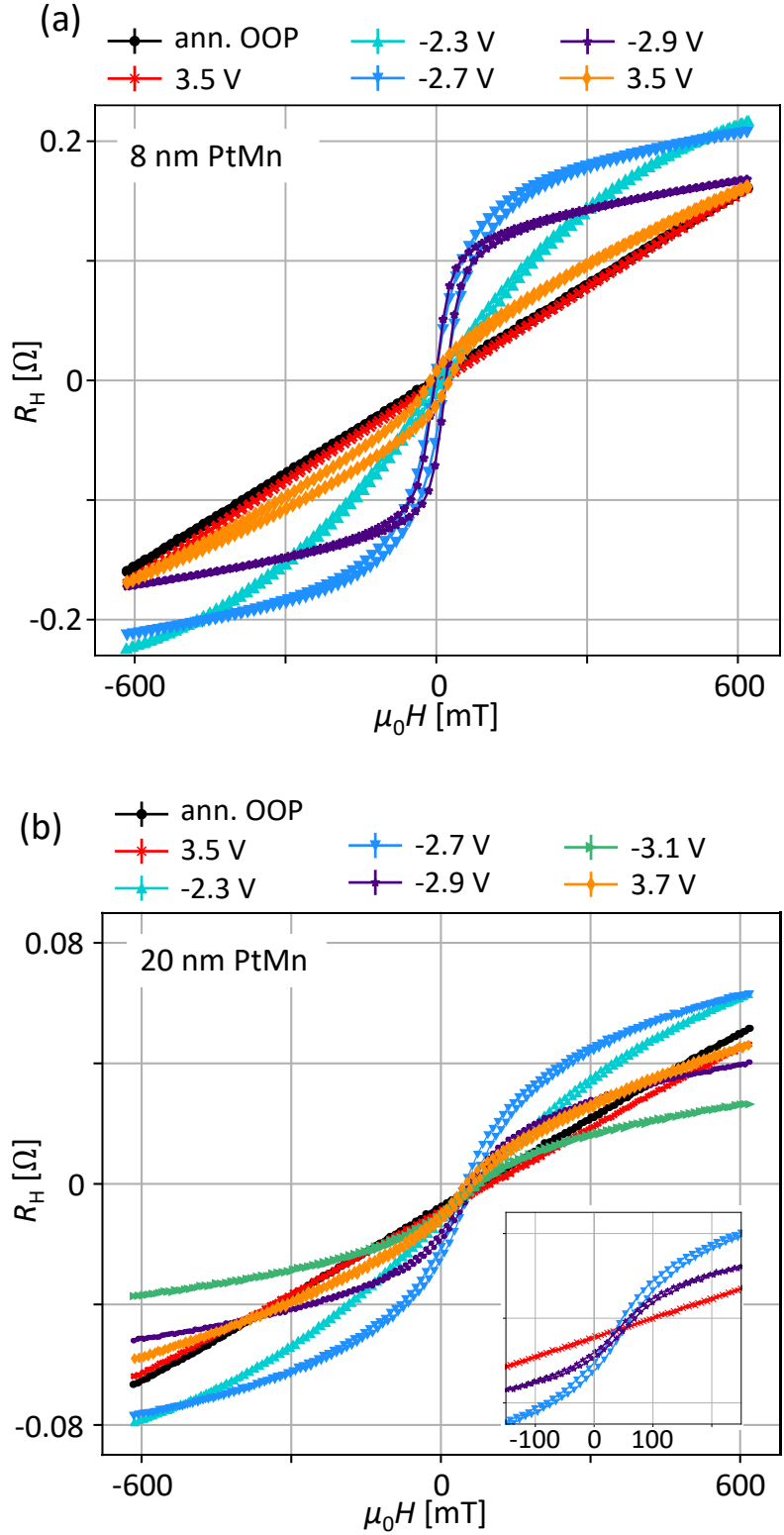

**FIG. S6.** Enlarged AHE curves from the first gating cycle for the samples with (a) 8 nm PtMn and (b) 20 nm PtMn. They correspond to Fig. 2a and 3a from the main manuscript, respectively. For the sample with 20 nm PtMn, two additional magnetization curves are shown here, after the application of -2.3 V and -2.9 V, which were left out in the main manuscript for clarity of the figure. The inset in (b) shows a zoom of the AHE curves after 3.5 V, -2.7 V and -2.9 V gate voltage, to display the analogy in the onset of overoxidation at -2.7 V for both samples with 8 as well as 20 nm PtMn.

## SM 7: Calculation of shift and coercivity of the AHE data

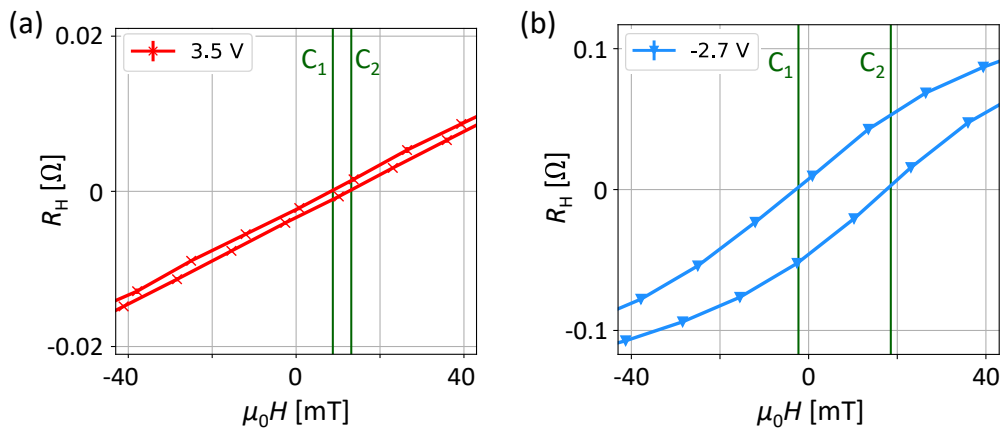

**FIG. S7.** *Determined coercive fields $C_1$ and $C_2$ displayed for two exemplary AHE curves of the sample with 8 nm PtMn (a) after the first application of 3.5 V and (b) after the application of -2.7 V gate voltage.*

The data analysis was done using Python. The magnetization curves were first interpolated with a step size of 1 mT. For both the upward as well as the downward half of the interpolated loops the coercive field was determined as the datapoint closest to zero resistance on the positive resistance side ($C_1$ and $C_2$ in Fig. S7). The shift and coercivity were calculated as the sum and difference of the two coercive fields, respectively.

The systematical error on the shift and coercivity was estimated to be 2.38 mT by investigating the shift of samples in the as deposited state. In that state, no exchange bias was set yet, so no shift should be present. Additionally, the error caused by the drift of the magnetization back towards the OOP state was added. The affected datapoints are the last two datapoints for the samples MASA697 D and E, after the application of 3.3 V and 3.5 V. The error was estimated as 4 mT and 5 mT, respectively, for the coercivity, and 1 mT for the shift. Before measuring the sample MASA697 F, the measurement program was updated, and the measurement speed significantly increased from 42 minutes to 11 minutes per measurement. Therefore, the data of MASA697 F was not significantly influenced by the slow drift anymore.

## SM 8: Reproducibility of the AHE measurements

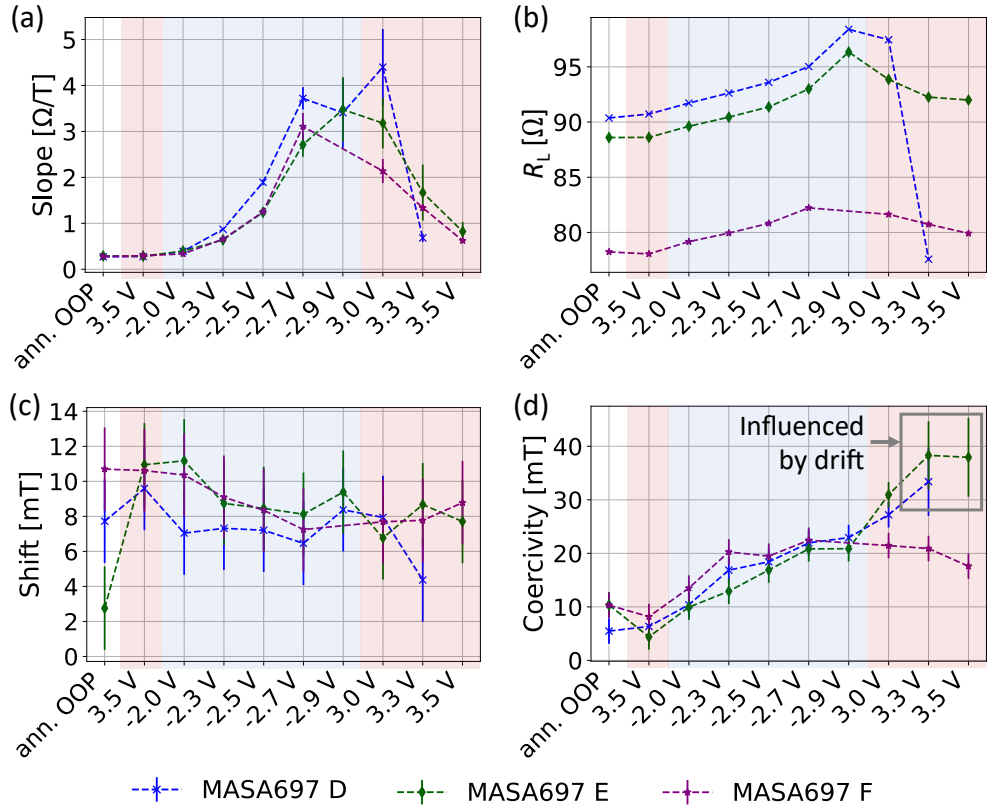

**FIG. S8.** *Overview over the slope, longitudinal resistance $R_L$, shift and coercivity, obtained from the AHE measurements of 3 nominally identical samples of Ta (5 nm)/Pt (3 nm)/PtMn (8 nm)/Co (0.9 nm)/HfO$_2$ (3 nm) with the labels MASA697 D, E and F. All samples were cut from the same wafer and annealed in the same run. The sample shown in the main text is MASA697 E. The grey box in (d) marks the datapoints which showed a significant drift during the measurement because of the evolution of the magnetic state back towards the OOP direction. The sample MASA697 F was not significantly influenced by this drift because the measurement speed was largely increased for that sample. The voltages on the x-axis represent the successively applied gating voltages and lines only serve as a guide to the eye.*

To check the reproducibility of the results measured via the AHE in the electromagnet, the results on three nominally identical samples were compared. The samples show qualitatively the same behavior.

The slope at zero remanence magnetization was determined as the average between the slope at the first and second zero-magnetization-crossing of the magnetization curve. The difference between the two values is given as the error bar. It is a measure for the asymmetry of the magnetization curve. In all samples, the slope shows a strong increase with negative gate voltages and a subsequent decrease with positive gate voltages. However, the samples show differences in the speed, in which they reach their maximum PMA state. For the green curve, the slope still increases from -2.7 V to -2.9 V, although we know from the SQUID measurements on this sample, that the total magnetic moment per area decreases after the application of -2.9 V. For the blue curve, the slope decreases at -2.9 V, so the decrease in the total magnetic moment has a larger influence on the slope than the increase in PMA. However, after the application of +3.0 V, when the magnetic moment partly recovered, the slope reaches a value larger than that for -2.7 V. This indicates that the speed in which the magnetic moment and the magnetic anisotropies change can slightly vary between the samples. The overall behavior remains unchanged in all three samples.

The plotted longitudinal resistance $R_L$ is the average longitudinal resistance at every gating step. The deviation of the datapoints at every gating step was used as the error. It is however smaller than the marker size. The longitudinal resistance shows an increase with a negative gate voltage, as expected from oxidation. With positive gate voltage, the resistance decreases again.

## SM 9: X-ray photoemission spectroscopy (XPS) information

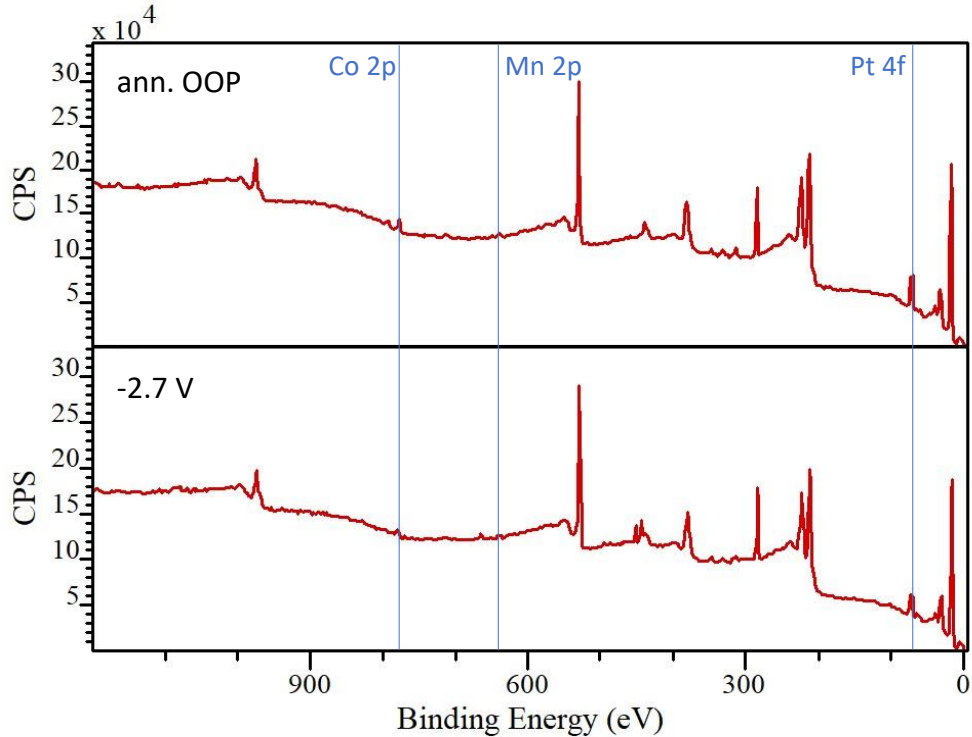

**FIG. S9.** XPS survey spectra for the sample with 8 nm PtMn, (a) annealed in an OOP field of 800 mT for 2 h at 300°C and (b) after applying a gate voltage of -2.7 V. The Co 2p, Mn 2p and Pt 4f peaks, marked by the blue lines, were additionally measured individually with a higher resolution. The Pt 4f peak was used for the charge compensation. The Co 2p and Mn 2p peaks are shown in Fig. 4 in the main manuscript.

XPS measurements were performed using a Kratos Axis Ultra DLD imaging photoelectron spectrometer. Measurements were carried out in the instrument's hybrid mode with a monochromated Al K$\alpha$ source (1486.7 eV) operating at 10 mA emission current and 15 kV voltage bias in an UHV chamber with a base pressure of approximately $1 \times 10^{-10}$ mbar. The source line width is approximately 1.0 eV FWHM, as calibrated with the Ag $3d_{5/2}$ line. The analysis area was 700 µm × 300 µm (i.e., X-ray spot size). The data was collected using a hemispherical analyzer at an angle of 0° to the surface normal. The takeoff angle was kept fixed along the surface normal in all experiments. Survey spectra (Fig. S9) were measured at a pass energy of 80 eV and the elemental spectra at 20 eV.

The analysis of the obtained spectra was done using the software 'casaXPS'. For the charge compensation the Pt 4f peak was aligned to its theory value, as Pt does not undergo oxidation. The background was subtracted for all Co spectra using the Spline Tougaard function[2]. For the Mn spectra, a linear background was subtracted because the background was rising towards smaller binding energies (see Fig. S9). All oxide peaks, as well as the satellite peaks, were fitted with the Gaussian/Lorentzian product formula 'GL(30)'. The asymmetric metallic peaks were fitted with a Lorentzian asymmetric lineshape 'LA(0.8,8,100)'. The theoretical peak positions, including multiplet splitting and satellites, were taken from the study by Biesinger et al.[3]